\documentclass{WileyMSP-template}
\usepackage{lineno}
\usepackage{siunitx}
\usepackage{makecell}
\usepackage{threeparttable}
\usepackage{ragged2e}
\usepackage{float} 
\usepackage{hyperref}
\usepackage{amsmath}

\begin{document}

\pagestyle{fancy}
%\rhead{\includegraphics[width=2.5cm]{vch-logo.png}}

\title{Cascaded Raman lasing in a lithium tetraborate (LB4) whispering gallery mode resonator}

\maketitle

% Author: Please give full first and last names for authors and include * after the name of all corresponding authors
\noindent\author{Chengcai Tian},
\author{Florian Sedlmeir},
\author{Jervee Punzalan},
\author{Petra Becker},
\author{Ladislav Bohatý},

\noindent\author{Keith C. Gordon},
\author{Richard Blaikie},
\author{Harald G. L. Schwefel*}

% Affiliations: Please provide adacemic titles (Prof. or Dr.) for all authors where applicable, and include an institutional email address for all corresponding authors
\begin{affiliations}
\noindent Chengcai Tian, Dr. Florian Sedlmeir, Prof. Richard Blaikie, A/Prof. Harald G. L. Schwefel\\
Department of Physics, University of Otago, Dunedin 9016, New Zealand\\
E-mail: harald.schwefel@otago.ac.nz

\noindent Chengcai Tian, Dr. Florian Sedlmeir, Jervee Punzalan, Prof. Keith C. Gordon, Prof. Richard Blaikie, A/Prof. Harald G. L. Schwefel\\
The Dodd-Walls Centre for Photonic and Quantum Technologies, New Zealand

\noindent Chengcai Tian, Jervee Punzalan, Prof. Keith Gordon, Prof. Richard Blaikie\\
The MacDiarmid Institute for Advanced Materials and Nanotechnology, Wellington 6012, NewZealand

\noindent Prof. Petra Becker, Prof. Ladislav Bohatý\\
Institute of Geology and Mineralogy, Sect.\ Crystallography, University of Cologne, 50674 Köln, Germany
\end{affiliations}

% Keywords: Please provide a minimum of three and a maximum of seven keywords, separated by commas


% Abstract should be written in the present tense and impersonal style (i.e., avoid we), and be at most 200 words long
\begin{abstract}

Lithium tetraborate (LB4) is a lithium borate compound and recently has shown renewed interest due to its exceptional linear and nonlinear optical properties. Its wide transparency range, spanning from \SI{0.16}{\micro m} to \SI{3.5}{\micro m}, and low loss in the visible range make LB4 highly popular in applications of harmonics generation and deep ultraviolet radiation. Also, LB4 is a good Raman-active material due to its high Raman gain. Here, a millimeter sized LB4 whispering gallery mode resonator (WGMR) is machined using single point diamond cutting, which has, to the best of our knowledge, the highest reported quality ($Q$) factor of $2.0 \times 10^9$ at \SI{517}{nm}. Then, stimulated Raman scattering (SRS) was investigated in this LB4 WGMR. When pumped with about \SI{7}{mW} at \SI{517}{nm}, four cascaded SRS peaks with wavelengths ranging from \SI{537}{nm} to \SI{608}{nm} are demonstrated, which can be clearly observed using an optical grating. Among them, the first order SRS is characterized and has a threshold of \SI{0.69}{mW} with a slope efficiency of 7.2\%. This is the first implementation of a LB4 whispering gallery mode Raman laser, which will facilitate usages of LB4 WGMR as compact Raman lasing source in future.

\end{abstract}

% Text: Please use section headings and subheadings as specified below. For communications, all section headings apart from Experimental Section should be removed
% Please make the first reference to a display item bold: \textbf{Figure 1}
% Do not abbreviate Figure, Equation, etc.; display items are always singular, i.e., Figure 1 and 2.
% Equations are always singular, i.e., Equation 1 and 2, and should be inserted using the {equation} environment, not as graphics
% Please do not use footnotes in the text, additional information can be added to the Reference list.

\section{Introduction}

Non-centrosymmetric borate crystals, with absorption edges larger than \SI{6.2}{eV} ($<$\SI{200}{nm}), such as LiB$_3$O$_5$ (LBO) and $\beta$-BaB$_2$O$_4$ (BBO), are popular nonlinear optical crystals to generate and manipulate ultraviolet light \cite{tran2016deep}. Lithium tetraborate (Li$_2$B$_4$O$_7$, LB4), a member of the borate family, has an exceptionally high absorption edge at \SI{7.56}{eV} (\SI{160}{nm}) \cite{petrov1998vacuum}. This maks LB4 a promising nonlinear optical material for harmonics generation \cite{komatsu1997growth} from the visible regime into the deep ultraviolet regime \cite{petrov1998vacuum}. On the other end of the spectrum its transparency ranges up to \SI{3.5}{\micro\meter} \cite{kwon1994characteristics,sugawara1998linear}, which surpasses the infrared (IR) transparency of KH$_2$PO$_4$ (KDP, \SI{1.4}{\micro\meter}), BBO (\SI{2.6}{\micro\meter}) and LBO (\SI{3.2}{\micro\meter}) \cite{nikogosyan2006nonlinear}. Especially in the range below \SI{600}{nm}, the absorption coefficient remains consistently below \SI{0.3}{m^{-1}} \cite{takahashi2009reduction}, which corresponds to an absorption limited intrinsic quality ($Q$) factor of above $10^8$ in a resonator made out of this material~\cite{strekalov2016nonlinear} and therefore important for efficient nonlinear conversion in it. Additionally, this low absorption can help mitigate thermal effects in the crystal. On top of that, among the borate crystals~\cite{komatsu1997growth}, LB4 has the highest laser damage threshold of \SI{40}{GW/cm^2} and the lowest hygroscopicity. 

LB4 is a negative uniaxial crystal and belongs to tetragonal nonsymmorphic space group $C^{12}_{4v}$, with $a=b=9.479\text{\AA}$ and $c=10.286\text{\AA}$ of the unit cell \cite{krogh1962crystal,krogh1968refinement}. Larger boules of LB4 crystal up to \SI{10}{cm} in diameter with high quality and good refractive index homogeneity have been grown by the modified Bridgman technique \cite{tsutsui2000growth, tsutsui2001growth}. The second-order nonlinear coefficients $d_{31}$ and $d_{33}$ are \SI{0.5}{\pico\meter/V} and \SI{3}{\pico\meter/V}~\cite{kasprowicz2003elastic, furusawa1991second} respectively, comparable to those of other borate crystals and about one tenth of these of lithium niobate (LN) \cite{nikogosyan2006nonlinear}. Because it belongs to point group $4mm$, other non-zero nonlinear coefficients are $d_{32}$=$d_{15}$=
$d_{24}$=$d_{31}$ under Kleinman’s symmetry conditions~\cite{boyd2008nonlinear}. Moreover, the Pockels electro-optic coefficients have been thoroughly
investigated~\cite{boyd2008nonlinear, mendez1999pockel,bohty1989electrooptical}, where $r_{13}$=$r_{23}$=\SI{4.03}{\pico\meter/V}, $r_{33}$=\SI{3.96}{\pico\meter/V} and $r_{42}$=$r_{51}$=\SI{-0.11}{\pico\meter/V}. Some optic-related material parameters of crystaline LB4 are summarized in \textbf{Table \ref{tab:LB4}} below, and more detailed physical properties can be found in Ref.~\cite{nikogosyan2006nonlinear} and \cite{kaminskii2006stimulated}. 

\begin{table}[H]\centering
\renewcommand{\arraystretch}{1.5} % Increased line spacing
\begin{threeparttable}
     
    \caption{\bf Some optic-related material parameters of lithium tetraborate (LB4)}
    \begin{tabular}{c||c}
        \hline\hline
        \multicolumn{2}{c}{\textbf{Lithium tetraborate (LB4): negative uniaxial crystal}} \\
        \hline
        \textbf{Point Group} &  $4 mm$ \\
        
        \textbf{Sellmeier equations}\cite{sugawara1998linear}\tnote{1} & 
        \makecell{$n_o=2.56431+\frac{0.012337}{\lambda^2-0.013103}-0.019075\lambda^2$ \\$n_e=2.38651+\frac{0.010664}{\lambda^2-0.012878}-0.012813\lambda^2$}\\
        
        \textbf{Transparency range}\cite{petrov1998vacuum,sugawara1998linear} &  \SIrange[range-phrase=--]{0.16}{3.5}{\micro\meter}  \\
        
        \textbf{$\chi^{(2)}$ nonlinear coefficients}\cite{kasprowicz2003elastic, boyd2008nonlinear} & \makecell{$d_{32}$=$d_{15}$=$d_{24}$=$d_{31}$=\SI{0.5}{\pico\meter/V}\\
        $d_{33}$=\SI{3}{\pico\meter/V}} \\
        
        \textbf{Pockels electro-optic coefficients}\cite{boyd2008nonlinear, mendez1999pockel,bohty1989electrooptical} & \makecell{$r_{13}$=$r_{23}$=\SI{4.03}{\pico\meter/V},
        $r_{33}$=\SI{3.96}{\pico\meter/V}\\
        $r_{42}$=$r_{51}$=\SI{-0.11}{\pico\meter/V}} \\
        
        \textbf{Raman scattering}\cite{kaminskii2006stimulated} & \makecell{Raman shifts: \SI{720}{cm^{-1}}, \SI{160}{cm^{-1}} \\Raman gain: $>$\SI{1.8}{cm/GW} \\ Raman linewidth: \SIrange[range-phrase=--]{8}{10}{cm^{-1}} }\\

        \textbf{Brillouin scattering}\cite{takagi1992brillouin}\tnote{2} & Brillouin shifts: $\approx$\SI{22}{GHz} and $\approx$\SI{30}{GHz}\\

        \textbf{Laser damage threshold}\cite{komatsu1997growth} & \SI{40}{GW/cm^2} at \SI{1.064}{\micro\meter}\\

        %\textbf{Elasto-optic coefficients\cite{krupych2016photoelastic}} & %\makecell{$p_{11}$=0.083, $p_{12}$=$p_{13}$=0.227, $p_{31}$=0.209,\\
        %$p_{33}$=0.163,$p_{44}$=-0.0075,$p_{66}$=-0.066}\\

        \textbf{Piezo-optic coefficients }(10$^{-12}$Pa$^{-1}$)\cite{krupych2016photoelastic} & \makecell{$\pi_{11}=-0.307$, $\pi_{12}=0.769$, $\pi_{13}=3.878$, $\pi_{31}=1.150$,\\ $\pi_{33}=1.639$,$\pi_{44}=-0.131$,$\pi_{66}=-1.387$}\\

        \textbf{Piezoelectric coefficients}\cite{shiosaki1985elastic} & $d_{15}$=\SI{8.07}{\pico\meter/V}, $d_{33}$=\SI{19.4}{\pico\meter/V}, $d_{31}$=\SI{-2.58}{\pico\meter/V}\\
        
        \textbf{Temperature derivative of refractive indices }(10$^{-6}$K$^{-1}$)\cite{sugawara1998linear}\tnote{3}  & $\frac{dn_o}{dT}=1.2$, $\frac{dn_e}{dT}=3$\\

        \textbf{Relative dielectric constants}\cite{senyshyn2010low}\tnote{4} & $\epsilon_{11}=7.4$, $\epsilon_{33}=9.7$ \\ 

        \textbf{Thermal expansion coefficients ($10^{-6}K^{-1}$)}\cite{shiosaki1985elastic}\tnote{5} & $\alpha_{11}=11.1$, $\alpha_{33}=-3.7$\\

        \textbf{Mohs hardness}\cite{komatsu1997growth} & 6 \\
        
        \hline\hline
    \end{tabular}
    \begin{tablenotes}
        \item[1] $\lambda$: wavelength in \SI{}{\micro\meter}
        \item[2] \SI{22}{GHz}: transverse acoustic modes along [100] and longitudinal modes along [001]; \SI{30}{GHz}: longitudinal modes along [100]
        \item[3] Temperature range: 20-\SI{40}{\degreeCelsius} and wavelength in \SI{0.43584}{\micro\meter}-\SI{0.64385}{\micro\meter}; Smaller than those of other borate crystals, KDP and LN. 
        \item[4] Frequency range: \SIrange[range-phrase=--]{1}{480}{KHz}
        \item[5] Temperature range: \SIrange[range-phrase=--]{20}{35}{\degreeCelsius}
        
    \end{tablenotes}
    \label{tab:LB4}
    \end{threeparttable}
\end{table}

These properties make LB4 an interesting base material for high-$Q$ monolithic optical resonators, such as whispering gallery mode resonators(WGMRs). Surprisingly the paper by Fürst et al.~\cite{furst2015second} has been the only work reporting on a LB4 WGM resonator and using it for second harmonic generation (SHG) of blue light.
In this paper we want to further explore the feasibility of LB4 WGMRs by fabricating a high-$Q$ resonator, characterize it at different wavelengths and demonstrate efficient Raman lasing in it.

WGMRs crafted from low-loss dielectric materials exhibit exceptional light confinement near the surface through total internal reflection (TIR) \cite{oraevsky2002whispering, vahala2003optical}. This unique property allows for remarkably high $Q$ factors and compact mode volumes \cite{matsko2006optical} spanning the whole transparency range of the used material (since TIR is broadband). Such properties make them a versatile platform for a wide range of applications in nonlinear optics and quantum optics \cite{strekalov2016nonlinear}. In particular, the wide spanning resonance spectrum enables efficient all-resonant nonlinear conversion processes such as parametric down conversion \cite{furst2010low}, SHG \cite{furst2015second} or, very prominently, frequency comb generation \cite{kippenberg_microresonator-based_2011} to name only a few.

Another interesting application for WGMRs is low-threshold Raman lasing or stimulated Raman scattering (SRS) \cite{grudinin2007ultralow,leidinger2016strong}.
SRS, a third-order nonlinear process in which photons interact with vibrating molecules or the lattice structure of materials, results in the photons being shifted to a frequency that is offset from its original frequency by the vibrational frequency it scatters from. Since SRS was discovered in 1962 by Woodbury et al.\ \cite{woodbury1962ruby, eckhardt1962stimulated} who realised that the extra line in their Ruby laser was from the vibration frequency of NO$_2$ group in liquid nitrobenzene, SRS has been applied widely in the fields of chemical spectroscopy, and medical bio-molecular imaging \cite{prince2017stimulated} because of its sensitivity to molecular vibrations. SRS is also used widely in laser applications, not only as frequency shifter to extend existing lasers \cite{hill1976low, spillane2002ultralow, kippenberg2004theoretical, grudinin2008efficient} but also in cascaded Raman lasers \cite{grudinin2007ultralow, min2003compact,rong2008cascaded}.
A WGMR Raman laser can be directly made out of any Raman-active materials such as silica \cite{spillane2002ultralow, min2003compact, kippenberg2004theoretical}, silicon \cite{rong2008cascaded} and crystalline materials, like calcium fluorite (CaF$_{2}$) \cite{grudinin2007ultralow, grudinin2008efficient, gold2022high, bhadkamkar2024high}, lithium niobate \cite{leidinger2016strong,yu2020raman} and here LB4, without the need of another cavity. %Here, we investigate SRS in a LB4 WGMR under a pump of \SI{517}{\nano\meter}.

LB4 has four kinds of Raman-active lattice vibrational modes based on the BO$_3$ and BO$_4$ groups, namely 18A$_1$+19B$_1$+19B$_2$+39E \cite{paul1982raman, gorelik2003raman, elalaoui2005raman,furusawa1990raman,wan2014raman}, where the two strongest branches manifest at \SI{720}{cm^{-1}} and \SI{160}{cm^{-1}}, respectively. The Raman spectra of our LB4 samples also show two consistently strongest peaks (see the first section of supplementary). The one (\SI{720}{cm^{-1}}) is provided by the planar BO$_3$ triangle groups' $A_1$ normal mode, and has a linewidth between \SIrange{8}{10}{cm^{-1}} \cite{burak2006origin,kaminskii2006stimulated}, while the other (\SI{160}{cm^{-1}}) is also probably related to $A_1$ mode but difficult to be assigned accurately \cite{elalaoui2005raman,gorelik2003raman}. Moreover, the Raman gain of LB4 is about \SI{1.8}{cm/GW}, about two orders of magnitude larger than that of silica \cite{stolen1973raman}, which can decrease the SRS threshold~\cite{spillane2002ultralow,matsko2003cavity} and motivates us to explore the Raman lasing in LB4 WGMRs. 

In this Letter, we report the fabrication of a millimeter sized LB4 WGMR, which has, to the best of our knowledge, the highest reported $Q$ factor ($2\times10^9$ at 517nm) in LB4. Then We report on the first cascaded SRS in it under a pump of \SI{517}{\nano\meter}.

\section{Experimental methods}

\begin{figure}[H]
\centering
\includegraphics[width=0.8\linewidth]{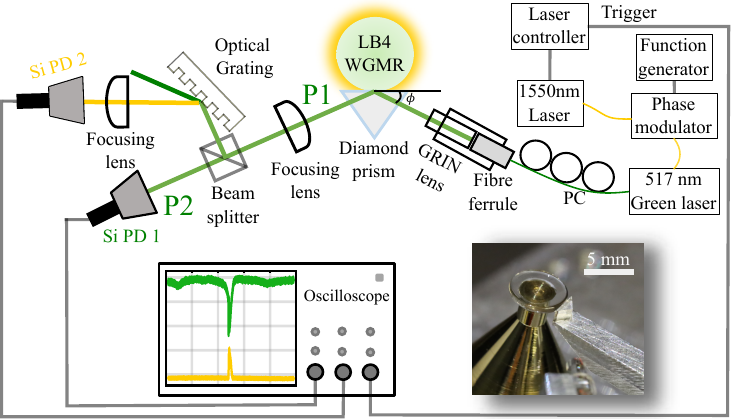}
\caption{Schematic of experimental setup. A green laser, generated using third harmonic of a telecom seed laser, goes through a phase modulator and a polarizaiton controller (PC) and is coupled into the LB4 WGMR via a fiber pigtailed ferrule, a graded-index (GRIN) lens and a diamond prism. A incoupling angle for the LB4 WGMR, denoted as $\phi=39^\circ$, is illustrated. The resonator and the prism are put on a rotation stage to adjust the incoupling angle. After the prism, both pump and signal modes are observed using two silicon photodiodes (Si PD 1 and Si PD 2) and an oscilloscope. An optical grating is used to spatially separate the generated cascaded signal from the pump. Green curves: pump; Yellow curves: Raman signal. The inset displays the LB4 WGMR with a major radius of \SI{2.97}{mm}$\pm$\SI{0.01}{mm} and a minor radius of \SI{2.30}{mm}$\pm$\SI{0.01}{mm}.}
\label{fig:setup}
\end{figure}

We fabricated our LB4 WGMR from a \SI{800}{\micro\meter}-thick (001)-orientated LB4 wafer grown by some of us \cite{kaminskii2006stimulated}. For fabrication we follow a now well established process for crystalline WGMRs (for a detailed description see for example \cite{grudinin_crystalline_2008} or \cite{sedlmeir2016crystalline}): A precursor disc is first drilled out of the wafer using a hollow brass drill and water-based \SI{30}{\micro\meter} diamond slurry. Subsequently, the disc was glued on a brass rod with hot wax and subsequently precisely shaped using single-point diamond turning \cite{sedlmeir2016crystalline}. We tested different rake angles between $-45^\circ$ to $-20^\circ$ and found that $-35^\circ$ yields the best surface quality after cutting (see the fifth section of supplementary). At this angle the cutting surface remained still rough with numerous cracks probably due to easy crack propagation on (100) and (010) planes of LB4 \cite{dolzhenkova2006crack}. Following the cutting process, the preliminary WGMR underwent meticulous manual polishing using diamond slurry with decreasing grain sizes from \SI{30}{\micro\meter} to \SI{0.25}{\micro\meter} until no scratches or other defects were observable under a microscope, a procedure that typically spanned several hours because of its high Mohs hardness. The resulting LB4 WGMR, as depicted in inset of \textbf{Figure~\ref{fig:setup}}, with azimuthal radius 
$R= \SI{2.97}{mm}\pm\SI{0.01}{mm}$ and polar radius $r= \SI{2.30}{mm}\pm\SI{0.01}{mm}$, was cleaned and put in the experimental setup. 

Our setup employed to characterize the linear properties of the resonator and the Raman lasing is sketched in Figure~\ref{fig:setup}. To stabilize the optical spectrum, the WGMR is placed on a temperature controller. A green laser at \SI{517}{\nano\meter} (EYLSA, Quantel), generated using third harmonic of a telecom seed laser, is directed via optical fibers to a polarization controller (PC). Evanescent coupling into the LB4 WGMR is achieved by focusing the laser through a Gradient-Index (GRIN) lens onto the inner surface of a diamond isosceles triangle prism under an angle of total internal reflection. The prism sits on a piezoelectric actuator to control the gap between the resonator and prism. Incoupling light beams are directed at an angle of $\phi=39^\circ$ with respect to the base of the prism near the WGMR to efficiently excite the lowest-order radial modes in the WGMR \cite{gorodetsky1999optical}, as illustrated in Figure~\ref{fig:setup}. The resonator and the prism are put on a rotation stage to adjust the incoupling angle. Subsequent to the diamond prism, the transmitted pump signal is collected using a silicon photodiode (Si PD 1, Thorlabs PDA36A2) to measure the mode spectrum of the resonator and calibrate the input laser power in the system. Prior to reaching the PD 1, a 50:50 beamsplitter directs a portion of the outcoupled pump and Raman signal to an optical grating, spatially separating the generated cascaded signal from the pump. After the optical grating, a second silicon photodiode (Si PD 2, Thorlabs PDA36A-EC) detects the intensity of different orders of the cascaded Raman lasing. The wavelength (and hence the order) of the cascaded Raman lasing signal was determined by replacing the PD 2 with an Ocean Optics spectrometer (Jaz). 

To measure the threshold of cascaded SRS accurately, the incident laser power in this system should be calibrated accurately. Here the power after the coupling prism (P1 shown in Figure\ref{fig:setup}) was firstly measured and calibrated (see the third section of supplementary). Considering the Fresnel reflections ($16.2\%$) for TM polarization at a diamond-air interface at the coupling angle, $\phi=39^\circ$, the incident power inside the prism is $\textrm{P0}=1.19\times \textrm{P1}$. The efficiency of the optical free space transmission pathway behind the prism is first calibrated when the diamond prism is positioned away from the LB4 WGMR. The calibrated efficiency, determined by comparing the power received by Si PD 1 (P2 shown in Figure\ref{fig:setup}) with the power just behind the prism (P1) measured using a power meter, is $\sim 0.40$. Then, the power after the prism (P1) for every measurement can be calculated using power on Si PD1 (P2) divided by this efficiency. Furthermore, the effect of coupling between the prism and the LB4 WGMR on the power P1 is investigated. When the prism and the LB4 WGMR are coupled, the measured P1 is approximately 3\% lower on average than that when they are not coupled, indicating the effect of the coupling between the resonator and the prism on the P1 is negligible (see the third section of supplementary).

\section{Results and Discussion}

\begin{figure}[H]
\centering
\includegraphics[width=0.8\linewidth]{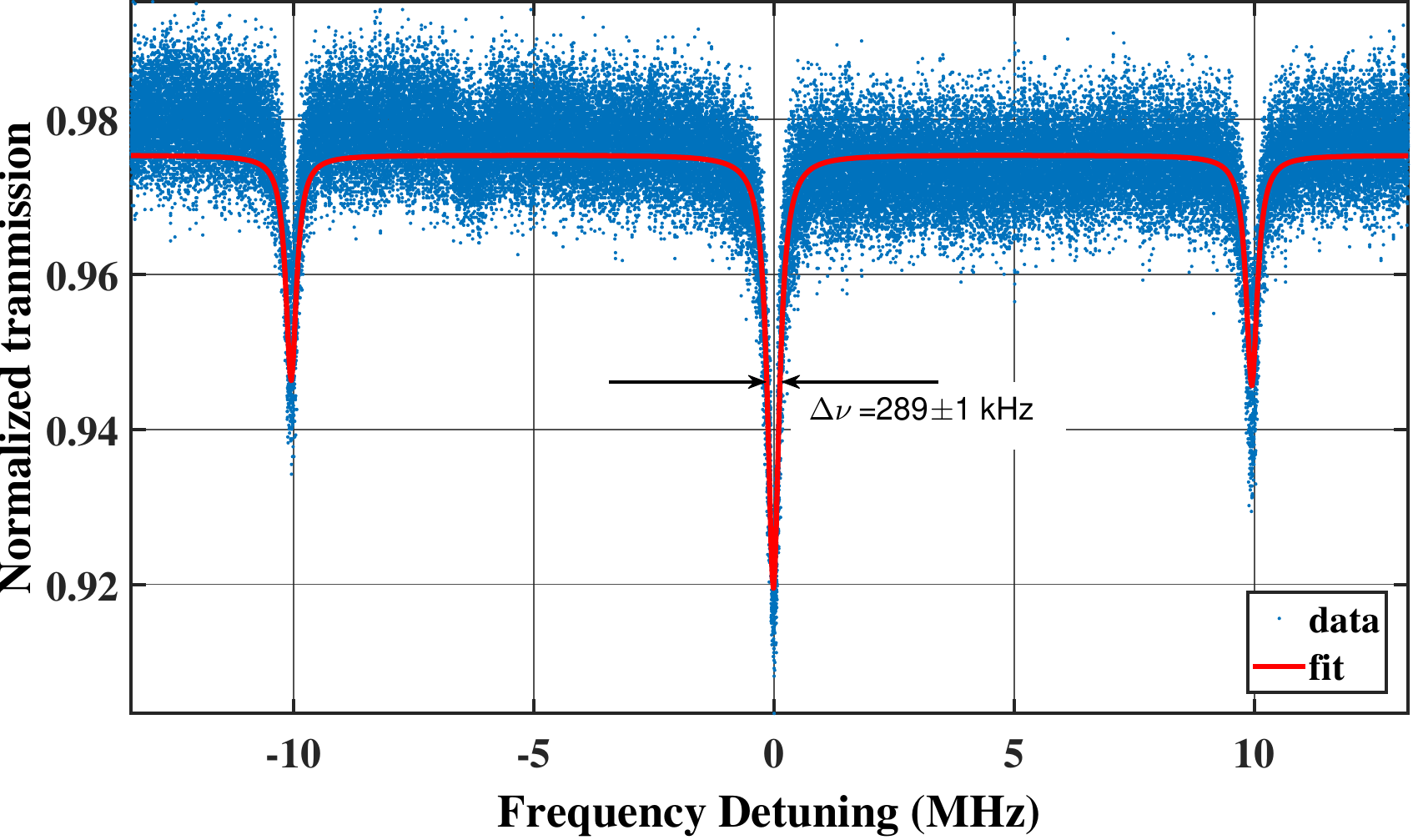}
\caption{One transverse magnetic (TM, electric field vertical to the symmetry axis of our WGMR) mode at under coupling state when the WGMR is excited by \SI{517}{\nano\meter}. The linewidth $\Delta \nu$ can be determined to be $289\pm1$ KHz from the Lorentzian fitting to an undercoupled mode with \SI{10}{M\hertz} sidebands, corresponding to $Q = 2.0 \times 10^9$. Blue dots: data acquired on the photodiode using the oscilloscope; Red curve: Lorentzian fitting.}
\label{fig:Q factor}
\end{figure}

The $Q$ factor and FSR of this LB4 WGMR are measured using a phase modulator with a function generator, which generates sidebands on both sides of the modes, as illustrated in \textbf{Figure \ref{fig:Q factor}} \cite{li2012sideband}. The linewidth, $\Delta \nu$, is determined by a Lorentzian fitting to a mode, then the $Q$ factor is subsequently determined from $Q=\nu/\Delta \nu$ ($\nu$: exciatation frequency). Using this modulation method, $Q$ factors are measured at \SI{517}{\nano\meter}, \SI{795}{\nano\meter} (DL100 diode laser, Toptica) and \SI{1550}{\nano\meter} (DLPro, Toptica)  for both transverse electric (TE, electric field parallel to symmetry axis of WGMR) and transverse magnetic (TM, electric field vertical to symmetry axis of WGMR) polarization. The best results of near-intrinsic $Q$ factor measurements (with the modes being strongly undercoupled during the measurement) are summarized in \textbf{Table \ref{tab:Q factor}}. 

\begin{table}[htbp]
\centering
\caption{\bf The best $Q$ factors of LB4 WGMR and inferred upper limit of absorption coefficients of LB4 at different wavelengths}
\renewcommand{\arraystretch}{1.5} % Increased line spacing
\begin{tabular}{cccc}
\hline\hline
\textbf{wavelength}&   \SI{517}{nm} & \SI{795}{nm}  &\SI{1550}{nm} \\
\hline
    \textbf{TM}&   $2.0\times 10^9$ & $1.2\times 10^9$ &  $6.1\times 10^7$\\
    \textbf{TE}&  $3.3\times 10^8$ & $1.0\times 10^9$ &  $6.8\times 10^7$\\ 
    \textbf{$\boldsymbol{\alpha}$ (m$^{-1}$)}&  0.010 & 0.011  &  0.095\\ 
    \hline\hline

\end{tabular}
  \label{tab:Q factor}
\end{table}

The $Q$ factor of mm sized WGM resonators is limited by either scattering due to remaining roughness of the surface or by intrinsic material absorption. Radiation losses are insignificant for this size and wavelength since they scale as $e^{-2m}$ \cite{garrett1961stimulated} where $m$ is the azimuthal mode number (number of field oscillations in one round trip of the light) which is on the order of 10000 in our resonator.
In any case, for the $Q$ factor, one can derive an upper limit from the material absorption of the used material using $Q_\text{intrinsic}=2\pi n/(\lambda \alpha)$ ($\lambda$: wavelength; $\alpha$: absorption coefficient of the materials).
For excitation at \SI{517}{nm} we find very high $Q$ factors of up to $2.0 \times 10^9$ for TM type modes (electric field polarized orthogonal to the resonator surface) and $3.3 \times 10^8$ for TE type modes (orthogonal to TM). The $Q$ factor of the TM modes corresponds to a material loss of $\alpha = \SI{0.01}{m^{-1}}$ which agrees well with the literature \cite{takahashi2009reduction}. The lower $Q$ factor of the TE modes might be a hint that bulk Rayleigh scattering is a significant contribution to the material loss at that wavelength: it was shown in \cite{gorodetsky_rayleigh_2000} that TM modes in WGM resonators can partly suppress bulk scattering effects leading to a higher $Q$ factor. 
At \SI{795}{nm} we find $Q$ factors of slightly above $10^9$ for both polarizations corresponding to a similar absorption coefficient of $\alpha = 0.011$ and at \SI{1550}{nm} the $Q$ factor is between $6 \text{ and } 7 \times 10^7$ corresponding to a significantly higher absorption coefficient of about $\alpha = 0.1 m^{-1}$ compared to the visible regime. The fact that $Q$ factors of both polarizations are similar at \SI{795}{nm} and \SI{1550}{nm} is also another evidence for the existence of the Rayleigh scattering at \SI{517}{nm}, because the scattering plays a smaller role at longer wavelengths. To the best of our knowledge absorption coefficients at \SI{795}{nm} and \SI{1550}{nm} have not been reported before for LB4, which, together with that at \SI{517}{nm}, are listed in Table \ref{tab:Q factor}.

In a WGMR, the SRS threshold is proportional to the mode volume and $1/Q^2$ \cite{matsko2003cavity}. Alongside the previously mentioned azimuthal mode number $m$, a WGM is characterized by the radial mode number $q$ and the polar mode number $p$ where $q$ is the number of field maxima along the radial direction and $p + 1$ the number of field maxima along polar direction. We want to identify the so called fundamental modes ($q = 1$, $p = 0$) since they have the smallest mode volume, consequently corresponding to the lowest SRS thresholds. The easiest way to do that is as follows \cite{schunk2014identifying}: first, the coupling angle is adjusted to the smallest value where a spectrum of the resonator can be obtained. The fundamental mode is part of the spectrum since it has the smallest propagation constant (corresponding to the smallest coupling angle). Higher order polar modes ($p > 0$) can be sorted out by their two lobe emission pattern. The $q$ number of the remaining modes can be determined by measuring their FSR - a larger FSR is linked to a higher $q$ number (compare second section of supplement). We find the fundamental mode has the smallest FSR of \SI{9.739}{GHz} and a coupling contrast of 32\%.

\section{Stimulated Raman scattering}

To investigate the conversion from the pump power to the Raman Stokes lines, we focus on the fundamental mode by decreasing the sweeping range of the laser around that mode, and gradually increase the pump power, while the pump on PD 1 and the signal on PD 2 are observed simultaneously. Upon reaching a certain pump power level, signal peaks appear on Si PD 2. Cascaded Raman lasing is confirmed by measuring the spectrum on the Ocean Optics spectrometer and a multimode optical fiber which replaces PD 2 in Figure~\ref{fig:setup}. The process is briefly described as follows: The first generated Raman-shifted photons are resonantly enhanced in the resonator and serve as the pump for the subsequent stages of Raman lasing. This process cascades and hence generates multiple higher-order Stokes lasing.

Cascaded SRS in a cavity was first observed in CCl$_4$ droplets using pulsed lasing by Qian and Chang in 1986~\cite{qian1986multiorder}, just one year after their initial observation of SRS in liquid droplets~\cite{snow1985stimulated}. Subsequently, in 2003, Min~\cite{min2003compact} first observed five cascaded SRS in a solid silica microsphere with a pump power of about \SI{0.9}{\milli \watt} and also applied coupled modes theory for its theoretical analysis. Since then, cascaded SRS has garnered significant attention and has been demonstrated in various types of resonators. For example, eight Stokes peaks were observed in CaF$_2$ resonators~\cite{grudinin2007ultralow}, two Stokes peaks in racetrack silicon micro rings~\cite{rong2008cascaded}, five Stokes peaks in As$_2$S$_3$ microspheres~\cite{vanier2014cascaded}, and two Stokes peaks in both aluminum nitride microrings~\cite{liu2017integrated} and silicon carbide microresonators~\cite{li2024efficient}, among others.

\begin{figure}[H]
\centering
\includegraphics[width=0.6\linewidth]{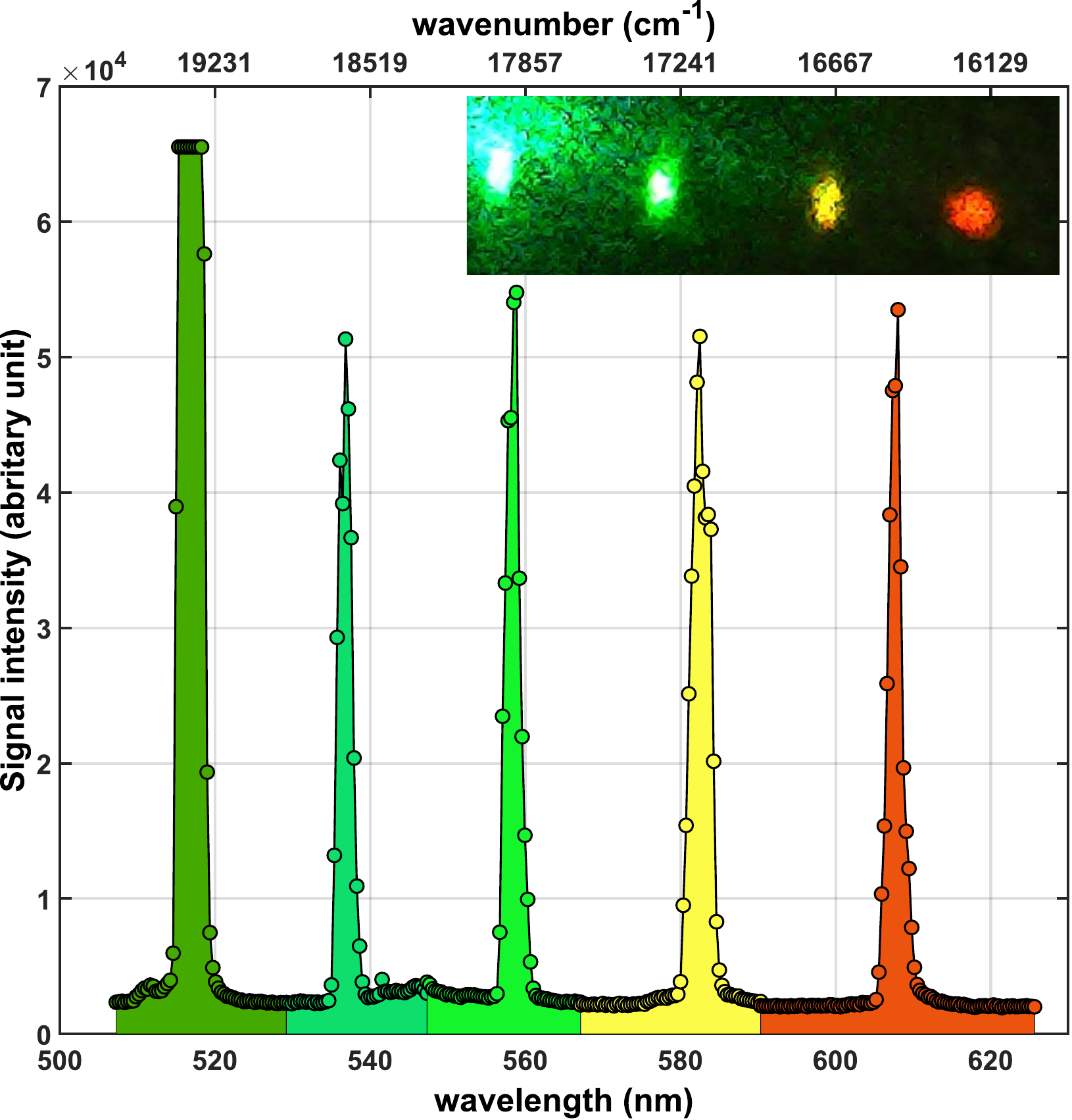}
\caption{Cascaded SRS recorded by OceanOptics in LB4 WGMR. The leftmost peak  is the pump laser at \SI{517}{nm}. These SRS peaks are located at \SI{537.1}{nm}, \SI{558.8}{nm}, \SI{582.4}{nm}, and \SI{608.2}{nm} respectively. Inset: A photograph of the cascaded SRS signal captured with an USB camera after the optical grating.}
\label{fig:cascaded SRS}
\end{figure}

In our LB4 WGMR, when the incoupled pump power is increased up to \SI{7}{mW}, cascaded SRS peaks with a wavenumber difference of about 720 cm$^{-1}$ up to the fourth order are observed using the Oceanoptics spectrometer, as shown in Figure \ref{fig:cascaded SRS}. The leftmost peak in Figure \ref{fig:cascaded SRS} is the pump laser at \SI{517}{nm}. These SRS peaks are located at \SI{537.1}{nm}, \SI{558.8}{nm}, \SI{582.4}{nm}, and \SI{608.2}{nm} respectively. The inset in Figure \ref{fig:cascaded SRS} shows a photograph of the cascaded SRS signal captured with an USB camera after the optical grating.

\begin{figure}[ht!]
\centering
\includegraphics[width=0.8\linewidth]{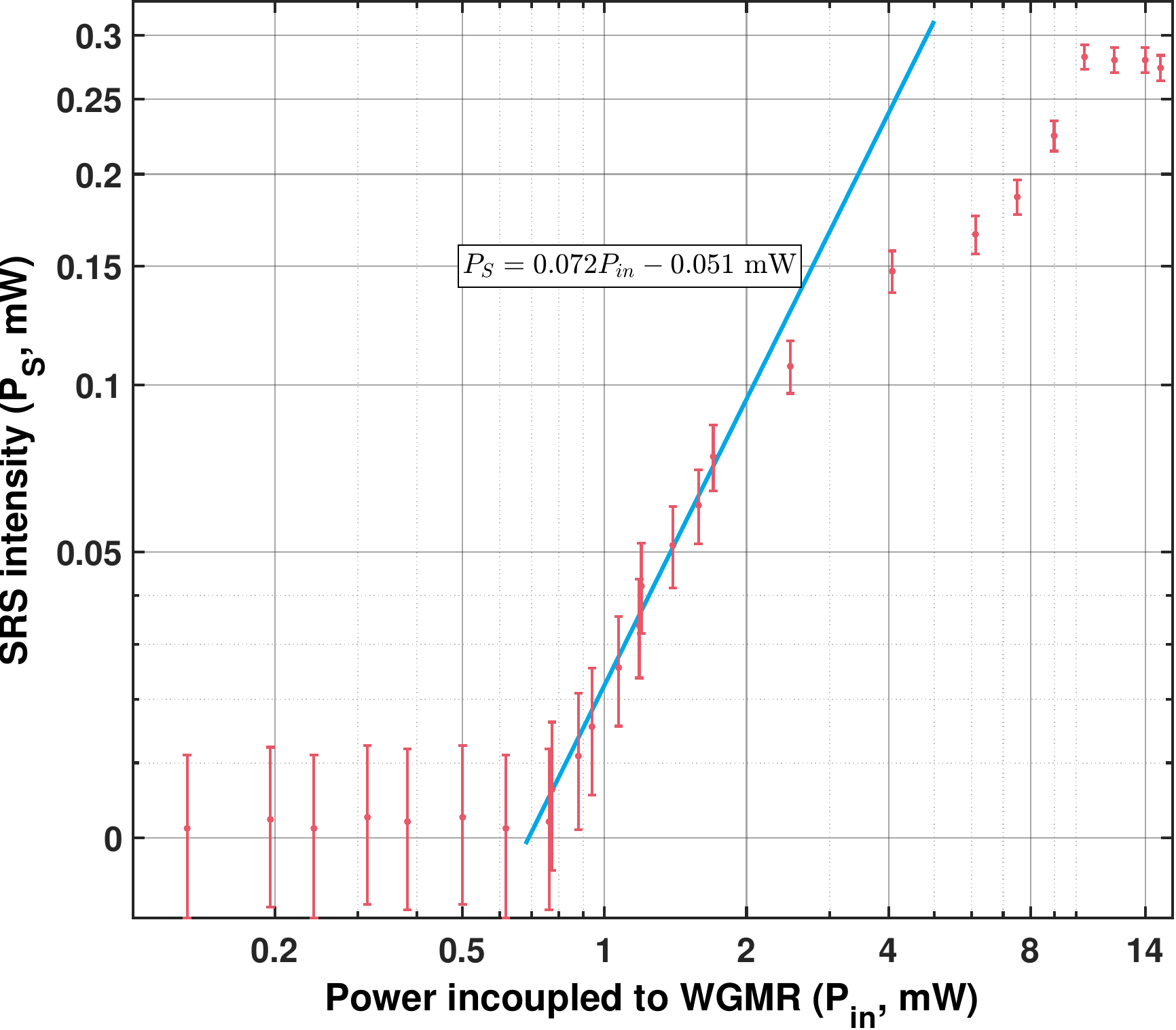}
\caption{First order SRS intensity changes with the power in fundamental mode (dots with error bars due to noise of photodiodes), the inserted solid line is the fit, from which the threshold P$_\text{{s}}=\SI{0.69}{mW}$, and the slope efficiency of SRS =\SI{0.072}{}.}
\label{fig:threshold}
\end{figure}

The first order SRS peak is identified using the spectrometer and its power is measured with the Si PD 2 at different incoupled powers. Notice here the incoupled power denotes the power coupled into the resonator, i.e., the product of mode contrast and the incident power inside the prism (P0). \textbf{Figure \ref{fig:threshold}} shows the power of the first Raman line versus the incoupled pump power P$_\text{in}$. The linear fit (until saturation) indicates a threshold power of \SI{0.69}{mW} for our LB4 WGMR. In comparison, the Raman lasing thresholds for other crystalline WGMRs stand at \SI{0.9}{mW} for LN WGMR~\cite{leidinger2016strong} and \SI{8.1}{mW} for magnesium fluoride (MgF$_2$) WGMR~\cite{tian2023magnesium}. To date, the lowest threshold in WGMR based Raman laser is \SI{3}{\micro W} in a CaF$_2$ WGMR with a significantly better $Q$ factor~\cite{grudinin2007ultralow}. 

As shown in Figure \ref{fig:threshold}, when the pump power exceeds the threshold of \SI{0.69}{mW}, the first-order SRS appears and its output power steadily increases with increasing pump power. Subsequently, as the coupled pump power is increased to \SI{2}{mW}, the first-order output power begins distorting and slowly increases because the intracavity power of first-order becomes sufficiently high to generate second order SRS \cite{kippenberg2004theoretical, grudinin2007ultralow,rong2008cascaded}. When the pump power is further increased to about \SI{7}{mW}, all four orders SRS appear and the output power of the first-order SRS stops increasing because those higher order SRS peaks deplete it. It is noteworthy that as the pump power continues to increase, there is a slight observable decrease in the first-order output power. This decline can be attributed to mode distortion induced by the thermal drift associated with highly elevated pump power levels \cite{carmon2004dynamical, grudinin2007ultralow}. Considering the reflection at each optical component and the diffraction efficiency of the optical grating, including the effect of the optical iris after the optical grating, which is used to block the pump signal also slightly affects the SRS signal, the observed unidirectional efficiency for the first-order SRS of fundamental mode is approximately 8.6\% according to the threshold curve, as detailed in the third section of the supplementary information.

\section{Conclusion}
In conclusion, we have demonstrated the highest reported $Q$ factor of $2.0 \times 10^9$ at \SI{517}{nm} in a LB4 WGMR. We also extracted absorption coefficients for longer wavelengths at \SI{795}{nm} and \SI{1550}{nm} which, to the best of our knowledge, have not yet been reported in literature. The high $Q$ factors in the visible regime which should, according to literature, extend into the UV down to \SI{250}{nm} and the $\chi^{(2)}$ nonlinear properties of LB4 make this crystal a very promising material for nonlinear optics in WGMRs.

We also observed strong Raman lasing with a very low threshold of \SI{0.69}{mW} which cascades as high as fourth order covering a spectral width from the pump at \SI{517}{nm} up to \SI{608}{nm}. The LB4 cascaded SRS can act as a compatible laser source for many applications in spectroscopy and like sciences. Moreover, the LB4 WGMR also has a $Q$ factor of $1 \times 10^9$ at \SI{795}{nm}, suggesting that the threshold for cascaded SRS to near infrared should be similar to that at \SI{517}{nm}, which can be widely applied in biomedical imaging and remote sensing.

\medskip
\textbf{Supporting Information} \par %Please delete the Suppporting Information statement if it is not applicable. Please supply Supporting Information in another file. Supporting information should not be provided in .tex format
Supporting Information is available.

\medskip
\textbf{Data availability} \par 
Data underlying the results presented in this paper are available after publishing at Cascaded Raman lasing in lithium tetraborate (LB4) WGMR (DOI: 10.5281/zenodo.10642630).

% Acknowledgements
\medskip
\textbf{Acknowledgements} \par %delete if not applicable))
The work is supported by Royal Society of New Zealand Marsden Council fund (contact 20UOO-080) and the University of Otago Postgraduate Publishing Bursary (Doctoral).

\medskip
\textbf{Conflict of Interest} \par
The authors declare no conflicts of interest.
% References
\medskip

% Use the following code if you wish to generate your bibliography with BibTeX;
% replace the string "MSP-template" below with the name(s) of
% the BibTeX data base(s) you want to use.
% The resulting bibliography-output (the content of the .bbl file)
% must be pasted back into this file before submission.
% Please also include your BibTeX data base file(s) in your submission
% so that we can re-run BibTeX if necessary.
%

\end{document}

\maketitle

\tableofcontents{}

%\author{} %leave this blank nb 
%% DO NOT ADD AUTHOR INFORMATION HERE; IT WILL BE ADDED DURING PRODUCTION

\begin{abstract}
Lithium tetraborate (LB4) is a lithium borate compound and recently has shown renewed interest due to its exceptional linear and nonlinear optical properties. Its wide transparency range, spanning from \SI{0.16}{\micro m} to \SI{3.5}{\micro m}, and low loss in the visible range make LB4 highly popular in applications of harmonics generation and deep ultraviolet radiation. Also, LB4 is a good Raman-active material due to its high Raman gain. Here, a millimeter sized LB4 whispering gallery mode resonator (WGMR) is machined using single point diamond cutting, which has, to the best of our knowledge, the highest reported quality ($Q$) factor of $2.0 \times 10^9$ at \SI{517}{nm}. Then, stimulated Raman scattering (SRS) was investigated in this LB4 WGMR. When pumped with about \SI{7}{mW} at \SI{517}{nm}, four cascaded SRS peaks with wavelengths ranging from \SI{537}{nm} to \SI{608}{nm} are demonstrated, which can be clearly observed using an optical grating. Among them, the first order SRS is characterized and has a threshold of \SI{0.69}{mW} with a slope efficiency of 7.2\%. This is the first implementation of a LB4 whispering gallery mode Raman laser, which will facilitate usages of LB4 WGMR as compact Raman lasing source in future.
\end{abstract}

\section{Raman Spectra of lithium tetraborate (LB4)}

\begin{figure}[ht!]
    \centering
    \includegraphics[width=\textwidth]{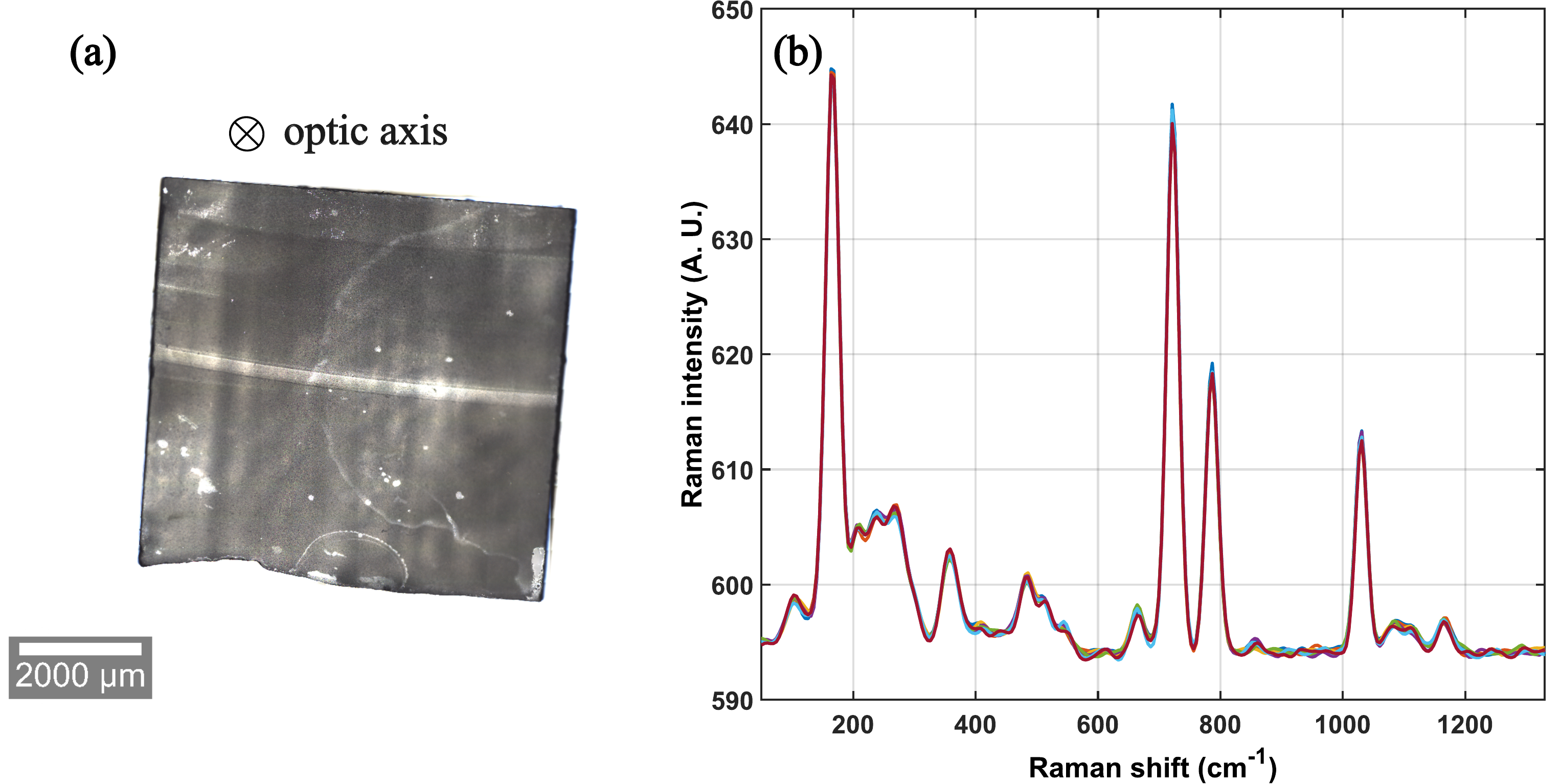}
    \caption{(a) The LB4 sample for Raman spectra measurement and its optic axis is perpendicular to the plane, shown by the black mark in the circle. (b) the multiple measurements of its Raman spectra at different positions and the highest intensity here are located at \SI{721}{cm^{-1}} and \SI{168}{cm^{-1}}.}
    \label{fig:LB4Raman_sepctra}
\end{figure}

To verify that the frequency shifts in our experiments equal the Raman shift of LB4, we measured Raman spectra of the LB4 sample we used to fabricate our LB4 whispering gallery mode resonator. The Raman spectra are measured using WITec Alpha300R confocal Raman microscope at the excitation of \SI{532}{nm}. Figure \ref{fig:LB4Raman_sepctra} (a) shows the LB4 sample for Raman spectra measurement. Its optic axis is perpendicular to the plane, shown by the black marker in the circle. Figure \ref{fig:LB4Raman_sepctra} (b) shows the multiple measurements of its Raman spectra at different positions and the highest intensity here are located at \SI{721}{cm^{-1}} and \SI{168}{cm^{-1}}, which is consistent with our measurement. In our resonator, we also observed Raman lasing peaks at about \SI{541.5}{nm} and \SI{545.8}{nm}, which respectively correspond to the cascaded Raman lasing with wavenumber difference about $5\times \SI{168}{cm^{-1}}$ and $6\times \SI{168}{cm^{-1}}$. Here, we don't choose to characterize \SI{168}{cm^{-1}} branched cascaded Raman lasing, because we find the first-order of this branch, due to smaller wavenumber difference, is masked by the strong pump, of which the threshold can't be characterized properly.

\section{Free spectral range measurement of the LB4 resonator}

\begin{figure}[ht!]
    \centering
    \includegraphics[width=\textwidth]{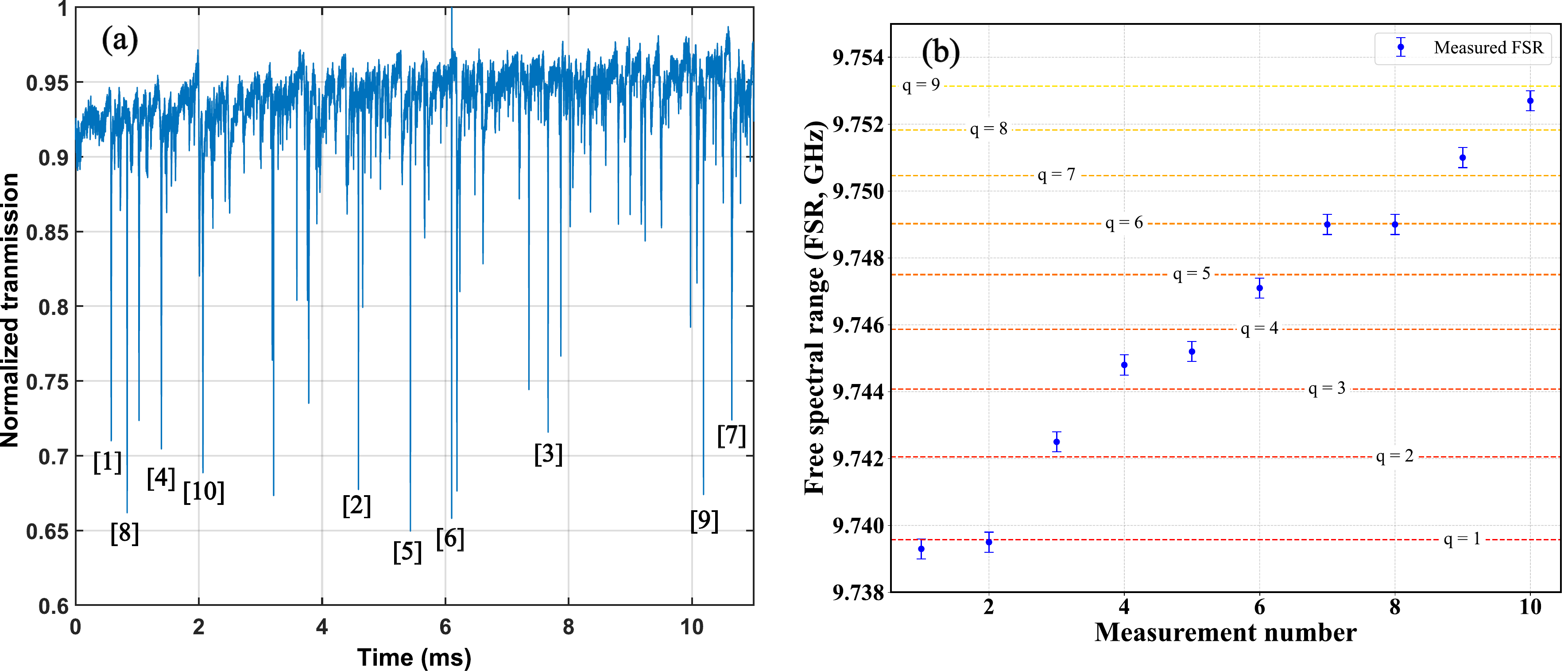}
    \caption{(a) The fraction of transmission spectrum of our LB4 WGMR within about \SI{5}{GHz} range and numbers in square brackets are corresponding to measurement numbers in (b). (b) Blue dots: ascendingly sorted measured FSRs of 10 modes with good contrast, error bars are from the linewidth of modes; Dashed horizontal lines are theoretical FSR values with corresponding $q$ at \SI{517}{nm}.}
    \label{fig:FSR}
\end{figure}

The fundamental mode ($q=1, p=0$) has the lowest thresholds due to the smallest mode volume. Therefore, to characterize the stimulated Raman lasing threshold, the fundamental mode needs to be identified \cite{schunk2014identifying}. Modes with polar mode number $p=0$ are typically identified as those with better coupling efficiency due to better spatial overlap between the beam coupled out of the resonator and the reflected pump from the prism \cite{sedlmeir2016crystalline}. The radial mode number $q$ changes the effective circulating length (or the effective refractive index) of modes, and therefore the free spectral range (FSR), which in turn helps us identify a mode's $q$ number. The FSR can be obtained by modulating frequency method \cite{li2012sideband}. When the laser is modulated with $\Omega \approx$ FSR, the sidebands of the $m+1$ and $m-1$ modes manifest symmetrically around the $m$ mode, and the frequency distance between sidebands and the $m$ mode is $\lvert \mathrm{FSR} - \Omega \rvert$. Initially, we estimate the major radius of our LB4 resonator to be ($R$) of \SI{2.97}{mm} $\pm$ \SI{0.01}{mm} with an optical microscope, allowing for a rough calculation of its FSR to be \SI{9.949}{GHz} using FSR=$c/2\pi Rn$ (with a bulk material refractive index $n=1.61$ at \SI{517}{nm}). Subsequently, we fine-tune the modulation frequency around this value for various modes until the sidebands align with those modes. At this point, the modulation frequency aligns with the FSR of this particular mode.

For this, light is first coupled into the WGMR to obtain its spectrum at the optimal coupling angle for TM polarized modes. Figure \ref{fig:FSR} (a) is the transmission spectrum of our LB4 WGMR within \SI{5}{GHz} range and numbers in square brackets are corresponding to measurement numbers in the Figure \ref{fig:FSR} (b). In Figure \ref{fig:FSR} (b), a measurement of 10 different modes in Figure \ref{fig:FSR} (a) of our LB4 WGMR is shown. The measured FSRs values (blue dots in Figure \ref{fig:FSR}) are arranged in ascending order, clearly indicating that the dispersion between different $q$ orders can be effectively resolved. The theoretical FSRs' predictions shown in Figure~\ref{fig:FSR} (dashed horizontal lines) are calculated from the dispersion equation for WGM resonators \cite{breunig2013whispering}. During the theoretical calculation, the radius of the resonator is changed slightly to \SI{2.97075}{mm} to match the smallest FSR (9.739 GHz $\pm$ \SI{0.0003}{GHz}) for $q=1$. There are two modes labelled ``1" and ``2", having the smallest FSR of \SI{9.739}{GHz}. Here, the mode labelled ``2", with a better coupling contrast of 32\%, is deemed as the fundamental mode because the GRIN lens (see Figure 1 in the main text) was set to be at the equatorial plane of the resonator and the beam shape is Gaussian, which make the fundamental mode couple better. Therefore, the mode labelled ``2" is selected to investigate SRS in our experiments. Furthermore, we find some higher order $q$ modes do not follow the theoretical FSRs perfectly which can be due to a number of factors, such as coupling to other modes and effect of higher-$p$ order~\cite{sedlmeir2016crystalline}. The cases $p = 0$ and $p > 0$ can be discriminated by their far-field emission pattern~\cite{schunk2014identifying,sedlmeir2016crystalline}. Specifically, modes with $p = 0$ exhibit a single-lobe emission pattern, while those with $p > 0$ display a two-lobe pattern.

\section{Power efficiency calculation of experimental setup}

\begin{figure}[ht!]
    \centering
    \includegraphics[width=0.8\textwidth]{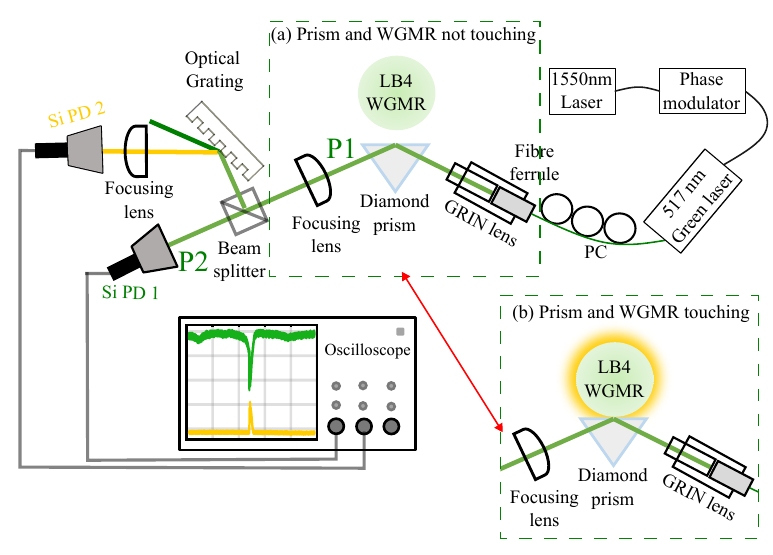}
    \caption{Schematic indicating the power calibration in our experiment. A green laser goes through a phase modulator and a polarizaiton controller (PC) and is coupled into the LB4 WGMR via a fiber pigtailed ferrule, a graded-index (GRIN) lens and a diamond prism. After the prism, the power P1 is measured using a power meter. The power P2 is derived from the silicon photodiode (Si PD 1) and an oscilloscope.}
    \label{fig:powerefficiency}
\end{figure}

To measure the threshold of SRS accurately, the incident laser power in this system should be calibrated accurately. Here the power after the coupling prism (P1 shown in Figure\ref{fig:powerefficiency}) was firstly measured and calibrated and then considering the Fresnel reflections ($16.2\%$) for TM polarization at a diamond-air interface at the coupling angle, $\phi=39^\circ$, the incident power inside the prism is $\textrm{P0}=1.19\times \textrm{P1}$. The efficiency of the optical free space transmission pathway behind the prism is calibrated using a power meter and Si PD 1 as shown in Figure \ref{fig:powerefficiency}. First, when the diamond prism is not touching the LB4 WGMR, we use a power meter to measure the power just behind the prism (P1) and Si PD 1 to measure the transmitted power (P2). All measured power values for calibration are reported in Table \ref{tab:efficiency}. From it, the efficiency between the power (P2) received by the Si PD1 and the power just behind the prism (P1) is $\approx 0.40$ when not coupling (Figure \ref{fig:powerefficiency} (a)). The expected efficiency is $(1-0.04)^5\times 0.5=0.41$ when the reflection coefficient is calculate as 0.04 via Fresnel equations at each surface of the focusing lens, beam splitter and photodiode. Next, the effect of coupling on input power is investigated. When the prism is touching the resonator, corresponding to Figure \ref{fig:powerefficiency} (b), We find that the power detected by Si PD 1 (P2$'$, shown in Table \ref{tab:efficiency}) is about 3\% on average less than that without coupling, indicating the effect of coupling on the input power is negligible. Finally, the incoupled power i.e., the power coupled into the resonator, can be calculated by the product of modes contrast and the incident power inside the prism (P0).

\begin{table}[ht!]
    \centering
    \caption{Power measurement behind the prism. P$_{\mathrm{laser}}$ is the display on the laser panel. P1 is the power after the prism measured using a power meter. P2 is the power measured by the Si PD1 when LB4 WGMR and prism are not touching, corresponding to Figure \ref{fig:powerefficiency} (a), while P2$'$ is the power when prism and LB4 WGMR touching.}
    \begin{tabular}{ccccc}
     
    \hline\hline
      P$_{\mathrm{laser}}$ (mW)  & P1 (mW) &P2 (mW) &P2$'$ (mW) & Efficiency (P2/P1) \\
      \hline\hline
      30  & 4.5 & 1.84 & 1.80 &0.41 \\ 
      60  & 8.9 & 3.57 & 3.47 &0.40 \\ 
      90  & 13.3 & 5.25 & 5.14 &0.40 \\ 
      120  & 17.5 & 6.95 & 6.84 &0.40 \\
      150  & 21.8 & 8.66 & 8.47 &0.40 \\ 
      180  & 25.9 & 10.50 & 10.02 &0.41 \\ 
      210  &29.6 & 12.06 & 11.49 &0.41 \\ 
      240  & 33.7 & 13.55 & 12.98 &0.40 \\ 
      270  & 38.1 & 15.28 & 14.73 &0.40 \\ 
      300  & 41.3 & 16.56 & 16.03 &0.40 \\
    \hline
    
    \end{tabular}
    \label{tab:efficiency}
\end{table}

\begin{figure}[ht!]
    \centering
    \includegraphics[width=0.7\textwidth]{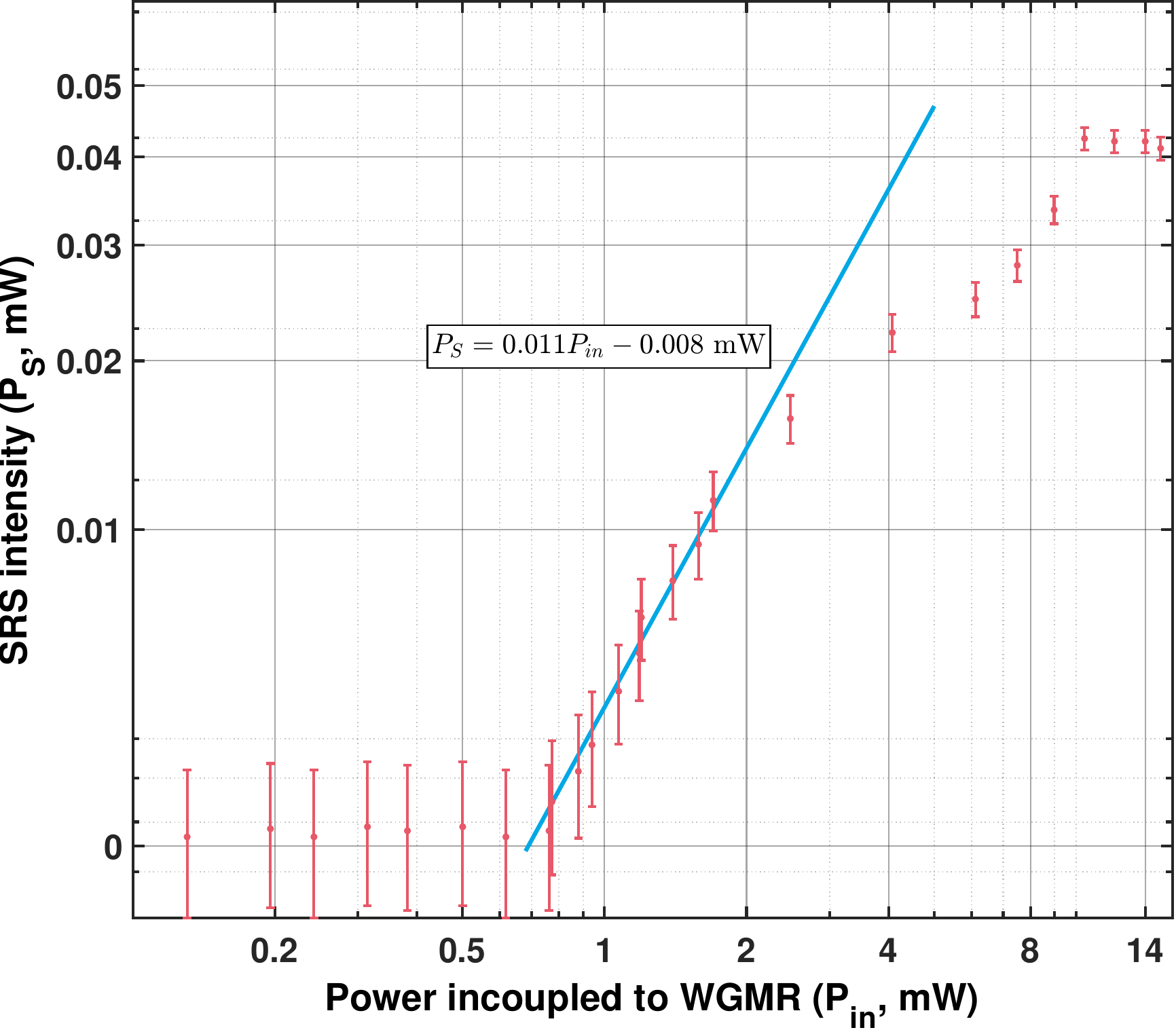}
    \caption{First order SRS intensity changes with the absorbed power in fundamental mode (dots with error bars due to noise of photodiodes), the inserted dashed line is the fitting curve when the pump is larger than the threshold P$_\mathrm{{ths}}$, of which the slope is the measured slope efficiency of SRS.}
    \label{fig:SRSthershold}
\end{figure}

For the Raman signal path, we use the same method above to calibrate. Considering the reflection coefficients of each optical components, beam splitter and diffraction efficiency of optical grating (Thorlabs GH13-24V, about 0.4 including the optical iris after the optical grating), the calculated efficiency is 0.15. Therefore the observed unidirectional efficiency for the first order SRS of fundamental mode is about 7.2\% from the measured slope efficiency (1.1\%) of threshold curve, shown in Figure \ref{fig:SRSthershold}.

\section{LB4 WGMR fabrication method}

\begin{figure}[ht!]
    \centering
    \includegraphics[width=\textwidth]{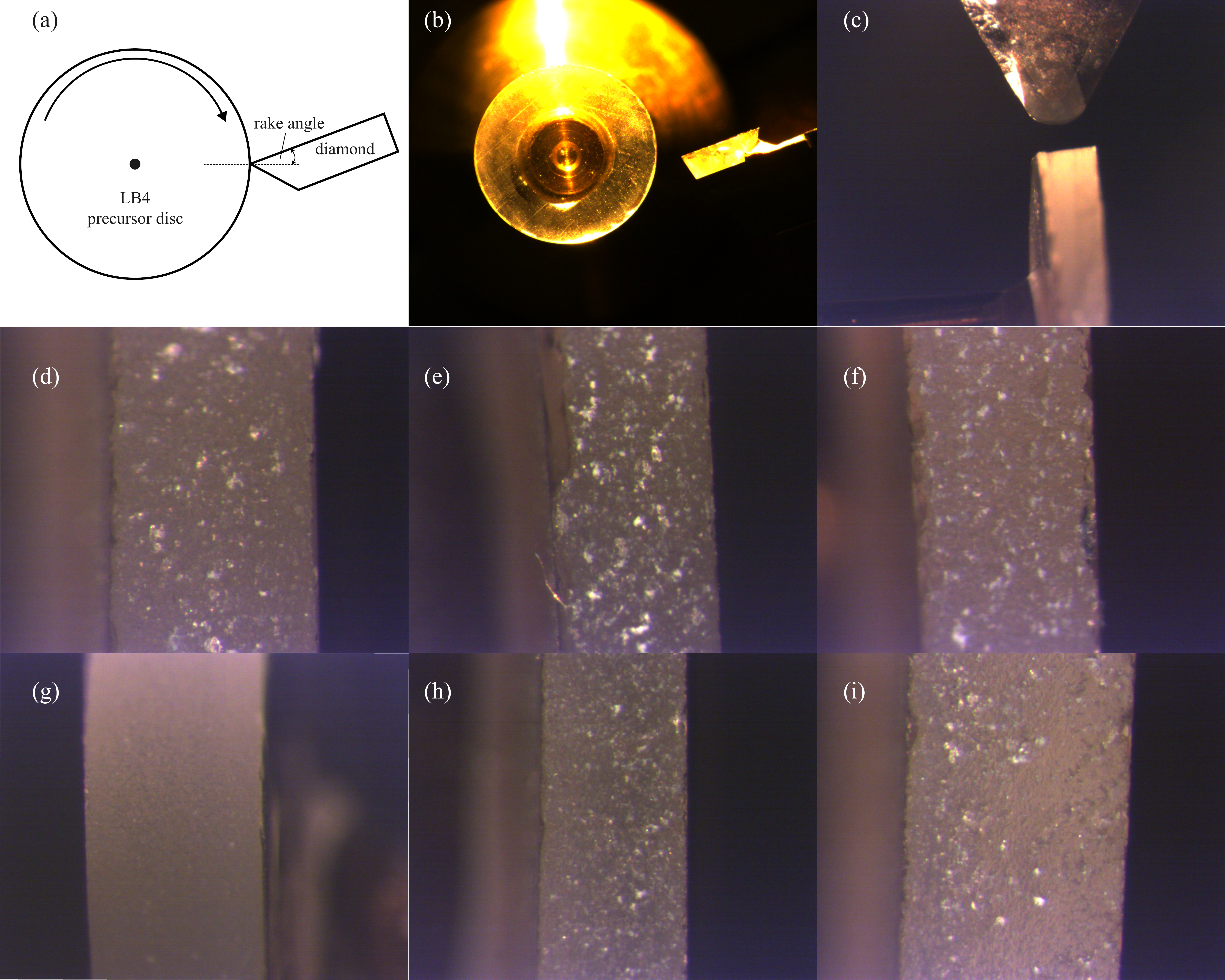}
    \caption{(a) Schematic of diamond cutting process, where the rake angle is depicted. (b) - (c) Side and top view during the cutting process. (d) - (i) Pictures of surface for different rake angle from $-20^\circ$ to $-45^\circ$ with a step size of $5^\circ$.}
    \label{fig:cutting}
\end{figure}

The LB4 WGMR used in this paper was machined by single-point diamond tuning from a (001)-oriented LB4 wafer with a thickness of \SI{800}{\micro\meter}. During the fabrication, the rake angle, depicted in Figure \ref{fig:cutting} (a), is an important parameter to optimize the surface quality. Figure \ref{fig:cutting} (b) and (c) show the side view and top view during the cutting process. Based on our previous experience on cutting brittle crystals, like lithium niobate (LN) and magnesium fluoride (MgF$_2$) \cite{sedlmeir2016crystalline}, we systematically explored various rake angles within the range of $-20^\circ$ to $-45^\circ$ with a step size of $5^\circ$, and pictures of surface just after each cutting are shown in Figure \ref{fig:cutting} (d) - (i). From surface roughness after cutting, we found the best rake angle for LB4 is $-35^\circ$. However, the cutting surface remained rough with numerous cracks probably due to easy crack propagation on (100) and (010) planes of LB4 \cite{dolzhenkova2006crack}. Thus, it still needs thorough manual polishing using diamond slurry with decreasing grain sizes from \SI{30}{\micro\meter} to \SI{0.25}{\micro\meter}.

